\documentclass[10pt,a4paper]{article}
\usepackage{graphics}
\usepackage{latexsym}
\usepackage{amssymb}
\usepackage{bm}
\usepackage{cite}
\begin{document}

\title{Frequency-shift keying signal detection via short time
stochastic resonance}

\author{Fabing Duan (1), Derek Abbott (2)\\
\small (1) Institute of Complexity Science, Department of
Automation Engineering, Qingdao University, \\ \small Qingdao
266071, People's Republic of China\\ \small (2) Centre for
Biomedical Engineering (CBME) and School of Electrical \& Electronic Engineering, \\
\small The University of Adelaide, Adelaide, SA 5005, Australia}

\maketitle
\begin{abstract}
A series of short time stochastic resonance (SR) phenomena,
      realized in a bistable receiver, can be utilized to convey
      train of information represented by frequency-shift keying (FSK)
      signals. It is demonstrated that the SR regions of the
      input noise intensity are adjacent for input periodic signals that differ
      in frequency appropriately. This establishes the possibility of
      decomposing M-ary FSK signals in bistable receivers. Furthermore,
      the mechanism of the M-ary FSK
      signal detection via short time SR effects is explicated in terms of the
      receiver response speed. The short time SR phenomenon
      might be of interest for neuronal information processing in non-stationary
      noisy environments, regardless
      of the short timescale or the frequency jitter of stimulus.
\end{abstract}

\section{\label{sec:I}Introduction}
Stochastic resonance (SR) is now a well established phenomenon
wherein the response of a nonlinear system to a subthreshold
periodic input signal can be enhanced by the assistance of noise
\cite{Benzi,Wiesenfeld,Bulsara1996,Gammaitoni1998,Moss,Harmer,Harmer2001}.
Since a single-frequency sinusoidal input conveys little
information content, this effect has been extended to aperiodic
(i.e. broadband) input signals, leading to the term: aperiodic
stochastic resonance (ASR)
\cite{Collins1995,Collins1996,Chapeau-Blondeau1997,Wiesenfeld1998,Goychuk2000,Barbay2000,Kish2001,Harmer1,McDonnell}.
From the point of view of information transmission, the aperiodic
information-bearing signal might be associated with analog
(amplitude and frequency) modulated signals
\cite{Anishchenko,Grigorenko,Morfu,Park,Comte}, or digitally
modulated signals
\cite{Chapeau-Blondeau1997,Barbay2000,Comte,Godivier,Duan,Xu,Mason}
within the context of SR effects. The digital pulse amplitude
modulated signal is intensively investigated for revealing new ASR
phenomena or novel applications
\cite{Chapeau-Blondeau1997,Barbay2000,Godivier,Duan,Xu,Mason}.
Recently, a new type of electronic receiver based on SR properties
has been proposed for retrieval of a subthreshold frequency-shift
keying (FSK) digitally modulated signals \cite{Morfu,Comte}.

In the present paper, the input information sequences are also
represented by equal-energy orthogonal signal waveforms that
differ in frequency, i.e. the FSK signals
\cite{Morfu,Comte,Proakis}. But, the prototype SR model, i.e. an
overdamped bistable system
\cite{Benzi,Wiesenfeld,Bulsara1996,Gammaitoni1998,Moss,Harmer,Collins1996,Chapeau-Blondeau1997,Wiesenfeld1998,Barbay2000,Grigorenko,Godivier,Duan,Xu,Asdi,Carusela},
is adopted as a nonlinear receiver that decomposes the received
signals. Usually, conventional SR or ASR characterizes the SR
phenomenon with a statistical measurement, resulting from a
long-term observational data
\cite{Benzi,Wiesenfeld,Bulsara1996,Gammaitoni1998,Moss,Harmer,Harmer2001,Collins1995,Collins1996,Chapeau-Blondeau1997,Wiesenfeld1998,Goychuk2000,Barbay2000,Kish2001,Harmer1,McDonnell}.
In contrast, we are more interested in the noise-enhanced effects
occurring in each short-term duration of each symbol interval as
the noise intensity increases, what we call the short time SR
phenomenon in this paper. The short time SR effect is consistent
with detecting weak signals from a short data record \cite{Asdi},
storing information in a short-term memory device \cite{Carusela}
and exploring transient stimulus-locked coordinated dynamics in
terms of mechanisms of short-term adaptation in sensory processing
\cite{Tass}. From a series of short time SR effects, the input
information contents can be deciphered at the output of the
bistable receiver in terms of different signal frequencies. The SR
effects realized in a nonlinear bistable receiver, as will be
shown, are not sensitive to adjacent periodic signals with an
appropriate frequency separation. In other words, the values of
the noise intensity at resonance regions or points are close for
two adjacent periodic signals. This nonlinear characteristic of a
bistable receiver is then studied for transmitting M-ary FSK
digitally modulated signals in detail. Furthermore, the response
speed of bistable receivers, independent of the frequency of input
signals, is theoretically deduced in Sec.~\ref{sec:III}. In view
of the receiver response speed, the mechanism of the M-ary FSK
signal detection is explained. Finally, we emphasize that the
short time SR is not a trivial effect in signal detection, for
example, in the detection of bipolar pulse signals with an unknown
arrival time \cite{Duan2}. The short time SR effect may also be of
interest in neurophysiology, even the stimulus exists in a
short-term timescale or has a frequency jitter. Neuronal noise, in
non-stationary noisy environments, might generically enhance the
prediction of the arrival information via the short time SR
effect, before the adaptive capabilities of neurons lowers their
thresholds to maximize information
transmission\cite{Zador,Stocks&Mannella}.

\section{\label{sec:II} Bistable receiver and M-ary FSK signal transmission}

\begin{figure}
\centering{\resizebox{8cm}{!}{\includegraphics{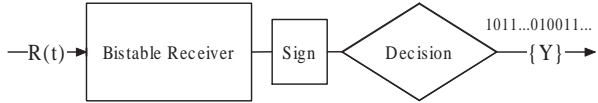}}}
\caption{Bistable receiver and demodulator for M-ary FSK signals.
\label{fig:one} }
\end{figure}

An input information-bearing sequence $\{I\}$ is mapped onto the
M-ary FSK signal $S(t)$ as
\begin{eqnarray}
S(t)=A \cos(2\pi f_m t), \, m=1, 2, \; \ldots, M, \label{eq:one}
\end{eqnarray}
for $(n-1)T\leq t\leq nT$, $n=1,2,\cdots$. Here, $A$ is the
amplitude, $\{f_m, m=1, 2, \ldots, M\}$ denotes the set of $M$
possible carrier frequencies corresponding to $M=2^k$ possible
$k$-bit symbols, and $T$ is the symbol interval. The received
signal $R(t)$ is
\begin{equation}
R(t)=A \cos(2\pi f_m t+\phi_m)+\eta(t),\label{eq:two}
\end{equation}
where $\phi_m$ are the phase shifts of carrier frequencies $f_m$
induced by the channel, and the background noise $\eta(t)$ is
additive Gaussian white noise with autocorrelation $\left\langle
\eta(t)\eta(0)\right\rangle =2D\delta (t)$ and zero-mean. Here,
$D$ denotes the noise intensity. Next, $R(t)$, as shown in
Fig.~\ref{fig:one}, is applied to a bistable dynamic receiver
given as
\begin{equation}
\tau_a\frac{dx(t)}{dt}=x(t)-\frac{x^3(t)}{X_b^2}+R(t),
\label{eq:three}
\end{equation}
with receiver parameters $\tau_a>0$ and $X_b>0$ \cite{Godivier}.
Here, $\tau_a$ is related to the system relaxation time. The
dynamics of Eq.~(\ref{eq:three}) is derived from the symmetrical
double-well potential $V_0(x)=-x^2/2+x^4/(4X_b^2)$, having the two
minima $V_0(\pm X_b)=-X_b^2/4$. Parameters $\tau_a$ and $X_b$ have
the units of time and signal amplitude respectively, and define
natural scales associated to the process of Eq.~(\ref{eq:three})
\cite{Duan}.

\begin{figure}
\begin{center}
\centering{\resizebox{7cm}{!}{\includegraphics{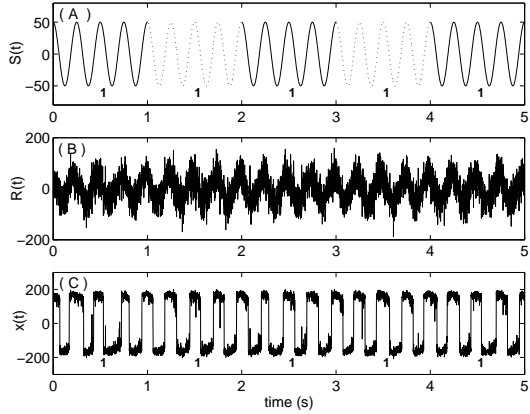}}}
\caption{Time evolution of binary FSK signal transmission in a
bistable receiver. (A) $S(t)=50\cos(8\pi t)$ representing binary
digit $1$ (or $0$). $S(t)$, plotted by solid and dotted curves
alternately in each symbol interval of $T=1$~s, should be
considered as segment signals rather than a continuous one in the
whole transmission time. (B) The received signal $R(t)$. Here, the
phase shift $\phi_m$ induced by the channel is $\pi/6$ and the
noise intensity $D=0.8$~V$^2$/Hz. (C) The bistable receiver output
signal $x(t)$ with parameters $\tau_a=1/3000$~s and $X_b=150$~V.
The decoded binary bits are depicted in terms of the zero crossing
times. The sampling time $\Delta t=10^{-3}$~s. \label{fig:two} }
\end{center}
\end{figure}

We are interested in recovering the successive input information
bits, from the observation of the system state $x(t)$. Moreover,
our focus is on the signal frequency rather than the waveforms at
the output of bistable receiver, due to the information contents
being represented by different frequencies. In this paper, we
numerically integrate the stochastic differential equation of
Eq.~(\ref{eq:three}) using a Euler-Maruyama discretization method
with a small sampling time step $\Delta t\ll\tau_a$ \cite{Gard}.
The demodulation method, as shown in Fig.~\ref{fig:one}, is
adopted with the zero crossing times $N_m$. If the bistable
receiver follows the subthreshold periodic signal correctly by the
assistance of noise, the zero crossing times $N_m$ of modulated
signal $S(t)$ should be $2f_mT$ in each symbol interval $T$. At
the output of the bistable receiver, however, the zero crossing
times $N_m$ will be in the vicinity of $2f_mT$ even in the
resonance region of noise, as shown in Fig.~\ref{fig:two}~(C). For
simplicity, the bytes represented by the corresponding input
signals with frequency $f_m$ are decoded as
\begin{eqnarray}
(f_{m-1}+f_m)T\leq N_m <(f_m+f_{m+1})T, \nonumber \\
or\;\; (2f_m-\Delta f)T\leq N_m <(2f_m+\Delta f)T, \label{eq:four}
\end{eqnarray}
with the frequency separation $\Delta f=f_{m+1}-f_m$ for
$m=1,\:2,\ldots$. Then, the output information sequence ${Y}$ is
decoded. Now, this system of Eq.~(\ref{eq:three}) with input
digits and output digital readings, can be viewed as an
information channel transmitting digital data. By comparing
sequences ${I}$ and ${Y}$, the measure of the percentage $P$ of
bytes correctly decoded will be used to quantify the performance
of this nonlinear information channel. We shall show that this
transmission of information can be assisted by additive noise
addition in each symbol interval, a property we interpret as a
short time SR effect.
\begin{figure}
\begin{center}
\centering{\resizebox{7cm}{!}{\includegraphics{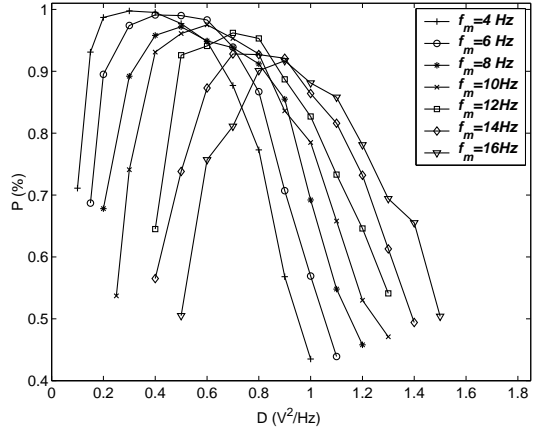}}}
\caption{The percentage $P$ of correctly decoded bytes as a
function of noise intensity $D$. In each symbol interval of $T$,
the input signal $S(t)=50\cos(2\pi f_m t)$ has frequency $f_m$ (as
marked in legends). The statistical values of $P$ are computed
numerically from $5000$ transmitted codes. The bistable receiver
is with parameters $\tau_a=1/3000$~s and $X_b=150$~V. The symbol
interval $T=1$~s and the sampling time $\Delta t=10^{-3}$~s.
\label{fig:three} }
\end{center}
\end{figure}
\begin{figure}
\begin{center}
\centering{\resizebox{6cm}{!}{\includegraphics{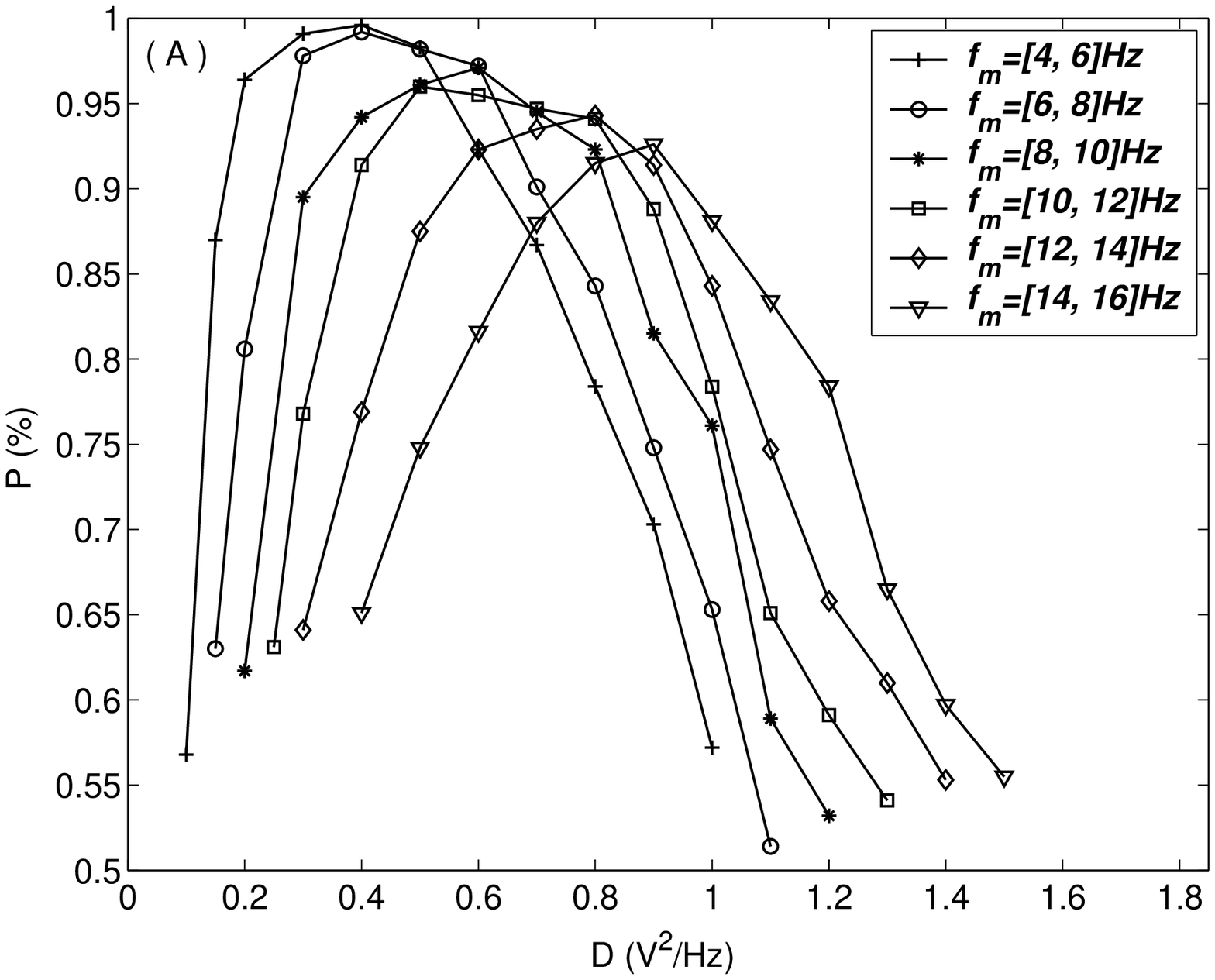}}}
\centering{\resizebox{6cm}{!}{\includegraphics{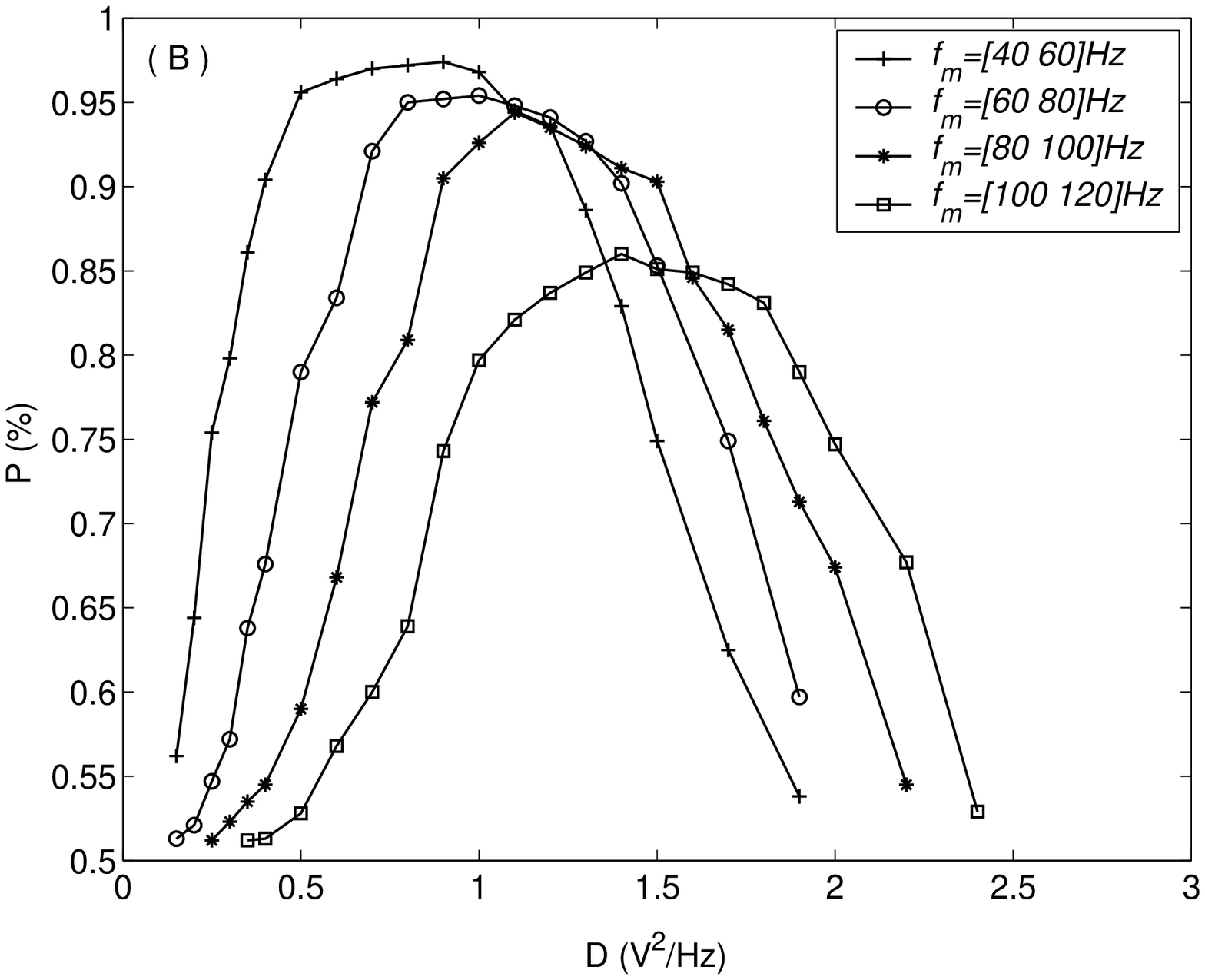}}}
\caption{Plots of the percentage $P$ as a function of $D$ for the
binary FSK signal transmission in bistable receivers. The
statistical values of $P$ are obtained from $5000$ transmitted
codes. The corresponding carrier frequencies are given in plots.
(A) The receiver parameters $\tau_a=1/3000$~s and $X_b=150$~V.
$T=1$~s and $S(t)=50\cos(2\pi f_mt)$.  (B) The receiver are with
parameters $\tau_a=1/10000$~s and $X_b=380$~V. $T=0.1$~s and
$S(t)=140\cos(2\pi f_mt)$. The sampling time $\Delta t=10^{-5}$~s.
\label{fig:four} }
\end{center}
\end{figure}
\begin{figure}
\begin{center}
\centering{\resizebox{6cm}{!}{\includegraphics{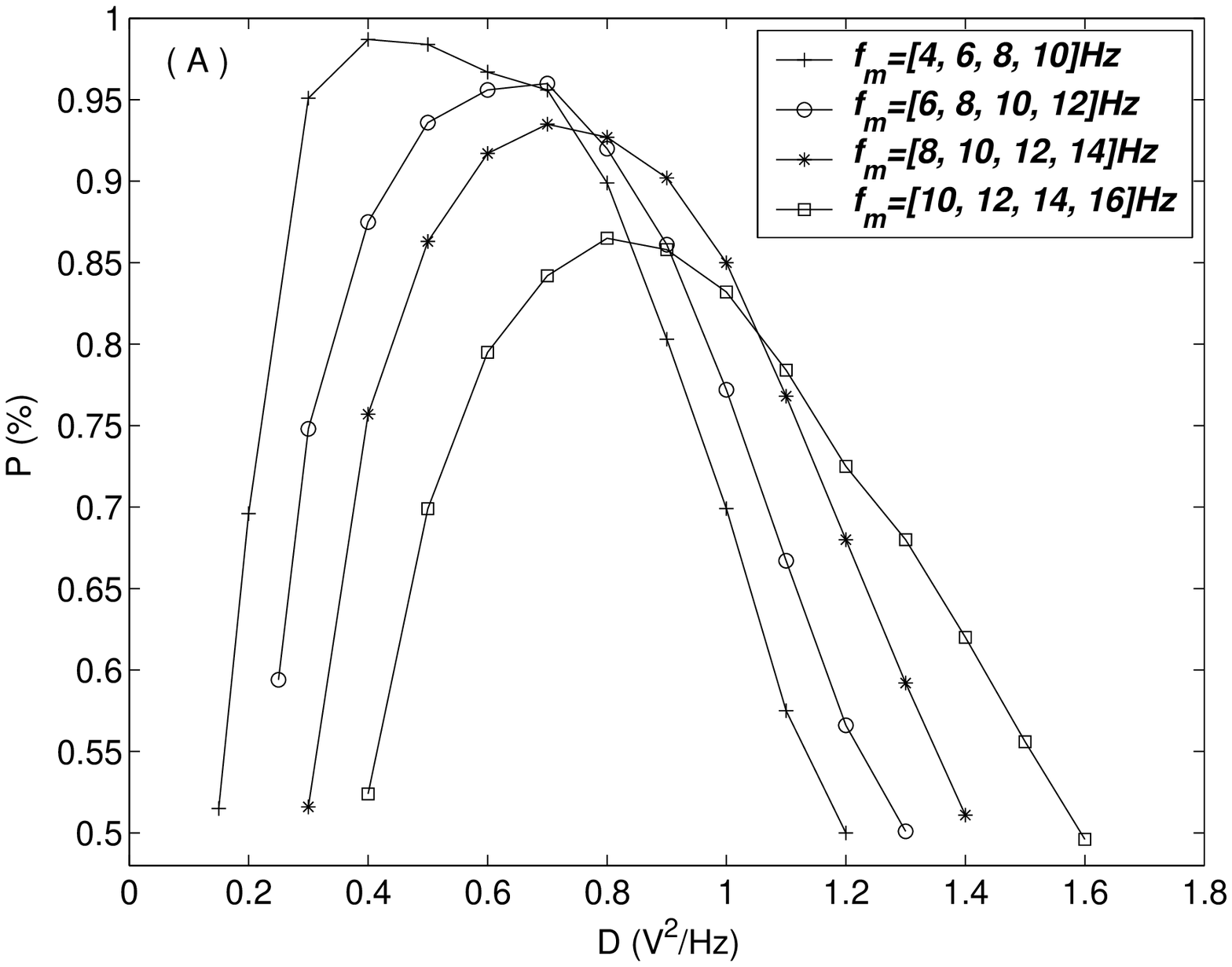}}}
\centering{\resizebox{5.8cm}{!}{\includegraphics{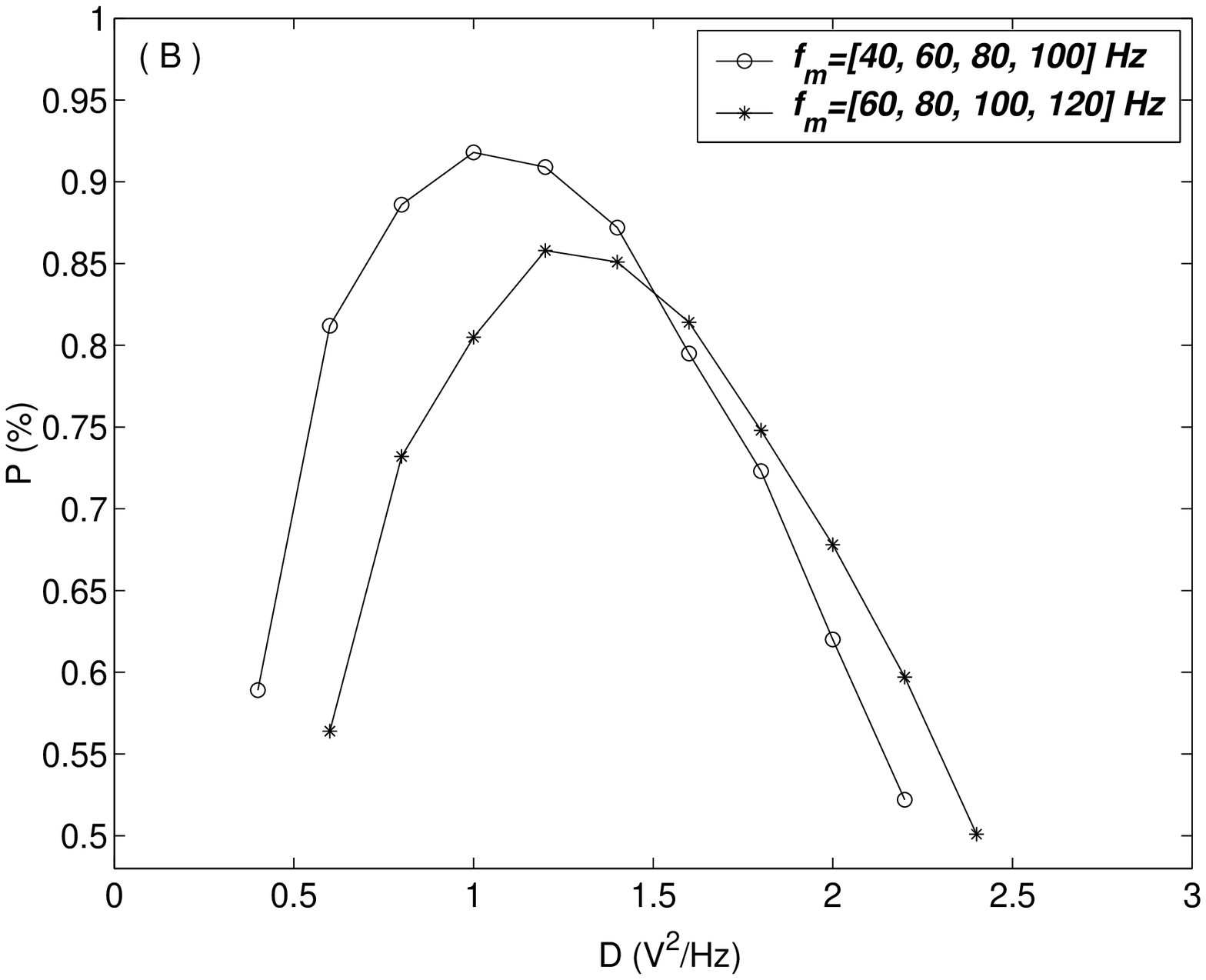}}}
\caption{Plots of the percentage $P$ as a function of $D$ for the
4-ary FSK signal transmission in bistable receivers. The
corresponding carrier frequencies are given in plots. (A) and (B)
are with the same parameters as in Fig.~\ref{fig:four},
respectively. \label{fig:five} }
\end{center}
\end{figure}

In Fig.~\ref{fig:two}, tuning the noise intensity $D$ so that the
switching between wells is made to agree closely with the input
periodic signal, resulting from the combined action of noise and
periodic signal in a bistable receiver. This synchronization
phenomena in each symbol interval of $T$ is called the short time
SR effect. In Fig.~\ref{fig:two}~(A), the input periodic signals
in each short time data record of $T$ have the same frequency,
which can represent a particular input binary sequence
${I}=[1\:1\:\cdots\:1]$ or ${I}=[0\:0\:\cdots\:0]$. The resonance
curves, illustrated in Fig.~\ref{fig:three}, are distinct from the
previous conventional SR form: the SR effects is explored in each
symbol interval $T$, rather than the entire transmission time,
recording the binary information. Numerical results of
Fig.~\ref{fig:three} show that the percentage $P$ of correct
switching events, as the noise intensity $D$ increases, presents a
typical SR characteristic. Especially worthy of note is that the
resonant regions or points of two adjacent frequencies are close.
This indicates that the bistable receiver is not sensitive to the
frequency jitter of input periodic signals. Thus, it is possible
to use a bistable receiver to decode binary or 4-ary information
represented by periodic signals with different frequencies,
i.e.~FSK digitally modulated signals. This possibility is
immediately demonstrated in Figs.~\ref{fig:four}
and~\ref{fig:five} with numerical experiments. In different symbol
intervals of $T$, the bistable receiver can follow the input
periodic signals with adjacent frequencies in an appropriate noise
intensity region. With the decision rule of Eq.~(\ref{eq:four}),
the percentage $P$ is a non-monotone function of noise intensity
$D$, resulting from a series of short time SR effects. For $M>4$,
the carrier frequency distance increases and a bistable receiver
is not sufficient for decoding the received FSK signals.
Naturally, a parallel bank of bistable receivers can be designed
for this complicated task, with different receiver parameters and
improved deciphered scheme.

\section{\label{sec:III} The mechanism of the FSK signal detection in a bistable receiver}

In this section we shall now attempt to explore the physical
mechanism of the FSK signal detection in a bistable receiver,
related to a an approximation of the nonstationary probability
density of Eq.~(\ref{eq:three}) and its temporal relaxation. The
temporal relaxation of nonstationary probability density, termed
the receiver response speed $\lambda_1$, will be demonstrated
being independence of the input signal frequency. The study of
system response speed allows us to explicitly have a deeper
understanding of the FSK signal transmission in a nonlinear
bistable receiver.
\begin{figure}
\begin{center}
\centering{\resizebox{7cm}{!}{\includegraphics{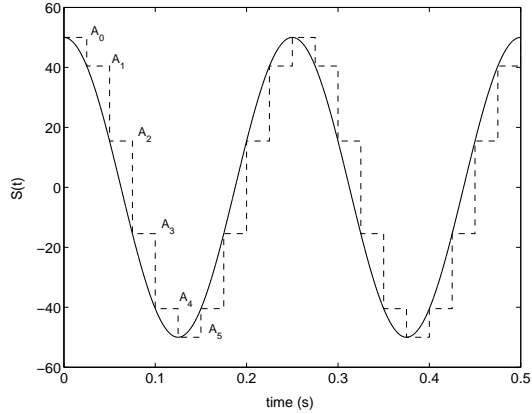}}}
\caption{A sinusoidal signal with the zero-hold sampling.
\label{fig:six} }
\end{center}
\end{figure}
\begin{figure}
\begin{center}
\centering{\resizebox{7cm}{!}{\includegraphics{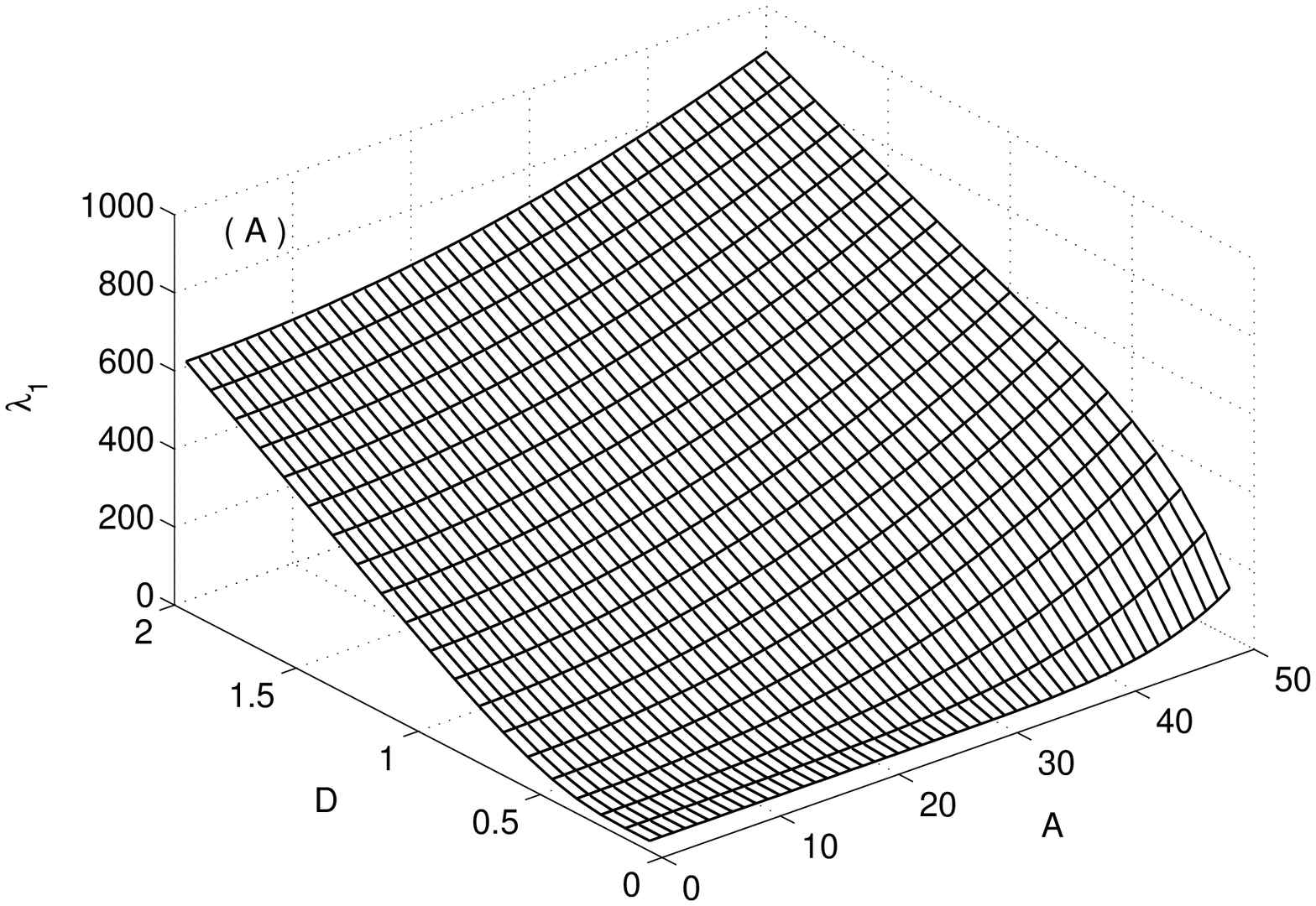}}}
\centering{\resizebox{7cm}{!}{\includegraphics{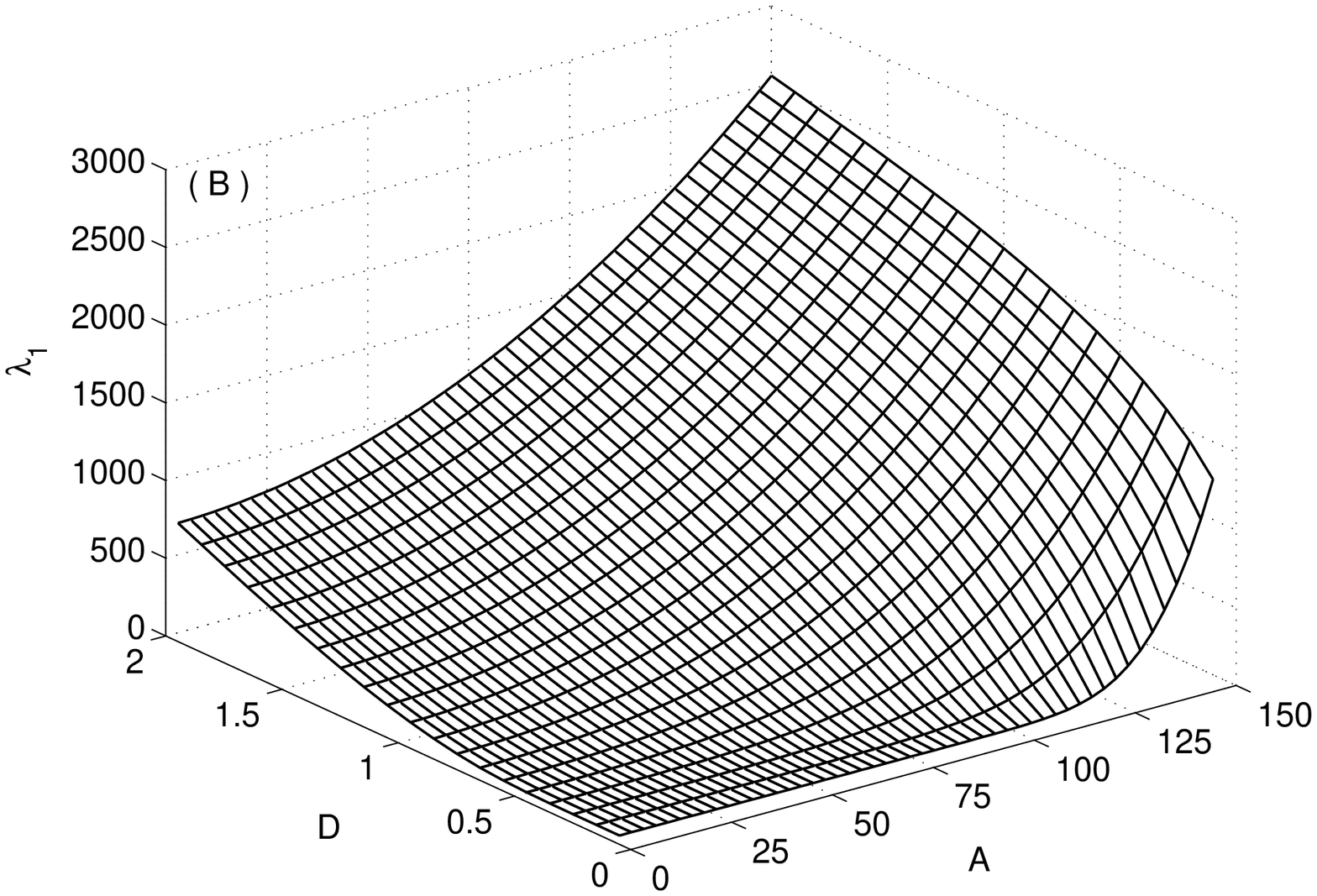}}}
\caption{The system response speed $\lambda_1$~(Hz) as a function
of signal amplitude $A$~(V) and noise intensity $D$~(V$^2/$Hz) for
the receiver with parameters (A) $\tau_a=1/3000$~s and $X_b=150$~V
and (B) parameters $\tau_a=1/10000$~s and $X_b=380$~V.
\label{fig:seven} }
\end{center}
\end{figure}

In numerical simulations, we numerically sample a sinusoidal
signal with zero-hold and sampling time $\Delta t\ll \tau_a$
\cite{Chapeau-Blondeau1997,Godivier,Duan,Xu,Duan2}. Hence, in a
sampling interval of $\Delta t$, the system of
Eq.~(\ref{eq:three}) is subjected to a constant amplitude
$s(t)=A_i$ ($i\Delta t\leq t < (i+1)\Delta t$, $i=0,1,2,\cdots$),
as shown in Fig.~\ref{fig:six}. In the presence of noise
$\eta(t)$, the statistically equivalent description for the
corresponding probability density $\rho(x,t)$ is governed by the
Fokker-Planck equation
\begin{equation}
\tau_a \frac{\partial \rho(x,t)}{\partial
t}=\big[\frac{\partial}{\partial x}
V'(x)+\frac{D}{\tau_a}\frac{\partial^2}{\partial
x^2}\big]\rho(x,t), \label{eq:five}
\end{equation}
where $V'(x)=-(x-x^3/X_b^2+A_i)$ and the Fokker-Planck operator
$L_{FP}=\frac{\partial}{\partial x}
V'(x)+\frac{D}{\tau_a}\frac{\partial^2}{\partial x^2}$.
$\rho(x,t)$ obeys the natural boundary conditions that it vanishes
at large $x$ for any $t$ \cite{Risken}. The steady-state solution
of Eq.~(\ref{eq:five}), for a constant input at $A_i$, is given by
\begin{equation}
\rho(x)=\lim_{t\rightarrow\infty}\rho(x,t)=C\exp[-\tau_a V(x)/D],
\label{eq:six}
\end{equation}
where $C$ is the normalization constant \cite{Risken}. As the
sampling amplitude $A_i$ varies, we encounter the nonstationary
solution $\rho(x, t)$ of the Fokker-Planck equation,
i.e.~Eq.~(\ref{eq:five}). This analysis is performed in
Appendix~A. We show in Appendix~A that the nonstationary solution
$\rho(x, t)$ can be expanded as an asymptotic representation by
eigenfunctions $u_i(x)$ and eigenvalues $\lambda_i$
\begin{equation}
\rho(x,t)=\sum_{i=0}^n C_i u_i(x)\exp[-\tau_a V(x)/(2D)]
\exp[-\lambda_i t], \label{eq:seven}
\end{equation}
where $C_i$ are normalization constants for $i=0,1,2,\ldots$.

\begin{figure}
\begin{center}
\centering{\resizebox{7cm}{!}{\includegraphics{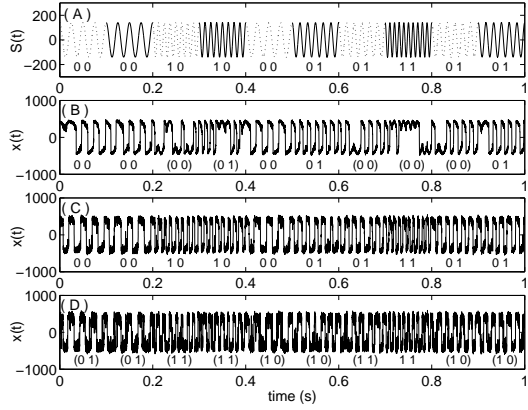}}}
\caption{(A) A 4-ary FSK signal $S(t)=140\cos(2\pi f_mt)$ with
carrier frequencies $40$~Hz, $60$~Hz, $80$~Hz and $100$~Hz,
representing 4-ary code words $00$, $01$, $10$ and $11$,
respectively. The receiver with parameters $\tau_a=10^{-4}$~s and
$X_b=380$~V. (B), (C) and (D) are plots of receiver outputs $x(t)$
at $D=0.25$~V$^2$/Hz, $D=1.0$~V$^2$/Hz and $D=2.4$~V$^2$/Hz. The
erroneous decoded bytes are bracketed. Here, $T=0.1$~s and $\Delta
t=10^{-5}$~s. \label{fig:eight} }
\end{center}
\end{figure}

Specifically, $\lambda_1$, termed the receiver response speed, is
a measure of the slowest time taken by the bistable receiver to
tend to the steady-state solution of Eq.~(\ref{eq:six}). In other
words, $\lambda_1$ is the speed of bistable receivers tracing the
variety of input signals, whence our term ``receiver response
speed.'' Figure~\ref{fig:seven} shows the behavior of the receiver
response speed in the related regions. Note that the receiver
response speed $\lambda_1$, for a fixed bistable receiver with
parameters $\tau_a$ and $X_b$, is a monotonically increasing
function of signal amplitude $A$ and noise intensity $D$, but
independent of the input signal frequency $f_m$, as indicated in
Eq.~(\ref{eq:A8}). When the noise intensity $D$ is too small, the
receiver cannot follow the input signal correctly, as shown in
Fig.~\ref{fig:eight}~(B). However, in the resonance regions of
noise intensity $D$, $\lambda_1$ is large enough to make the
receiver output $x(t)$ reach the steady-state, whereas the signal
amplitude continually changes from $\pm A$ to $\mp A$ at different
but adjacent carrier frequencies $f_m$. This indicates detecting
weak FSK modulated signals is possible in a bistable receiver, as
seen in Fig.~\ref{fig:eight}~(C). After the noise intensity $D$ is
beyond the resonance region, $\lambda_1$ is too fast to catch up
with the fluctuations induced by noise, resulting in the loss of
synchronization, as shown in Fig.~\ref{fig:eight}~(D).

The receiver response speed $\lambda_1$ contributes to the
mechanism of transmitting weak FSK modulated signals in a bistable
receiver: Within a reasonable region of noise intensity, the
receiver responds synchronically to the input periodic signal,
regardless of the short-term timescale of symbol interval $T$ or
the frequency jitter of signals (i.e.~the different carrier
frequencies $f_m$). Therefore, the SR effect realized in a symbol
duration $T$, i.e.~the short time SR, can be utilized to convey or
store the digital data \cite{Asdi,Carusela}. Additionally,
noise-enhanced frequency discrimination was reported in
Ref.~\cite{Zeng1}, and the frequency robust characteristic of
bistable model is also and verified in Ref.~\cite{Duan2} in
detail.

\section{\label{sec:IV} Discussion}
A detection strategy of subthreshold FSK digitally modulated
signals in a nonlinear bistable receiver is analyzed. It was
numerically demonstrated that the resonance values of noise
intensity appear closely for FSK modulated signals with adjacent
frequencies. Assuming a demodulation method of the zero crossing
times, a series of SR effects appearing in each symbol interval,
what we call the short time SR, provide the possibility of
detecting subthreshold M-ary FSK signals in bistable receivers. In
order to understand the mechanism of the FSK signal detection more
deeply, we introduce the receiver response speed, i.e.~a
theoretical measure for the receiver tending to a steady-state. In
the resonance region, the receiver response speed is fast enough
to trace the variety of input periodic signals, regardless of the
frequency difference and the short-term timescale of symbol
interval.

Finally, we argue that the short time SR effect might be an
optional detection strategy in neurodynamics. Neural noise will
play a positive beneficial role for information processing, in the
case of the stimulus exists in a short time duration or has a
jitter in frequency \cite{Zeng1,Asdi,Carusela}. Even under the
assumption that neurons have sufficient adaptability in sensory
systems\cite{Zador,Stocks&Mannella}, neural noise might provide an
enhancement via the short-time SR phenomenon before the neuron
adaptability acts. This is an open question and currently under
study.

\section*{Acknowledgements}
Funding from the Australian Research Council (ARC) is gratefully
acknowledged. This project is also sponsored by Scientific
Foundation for Returned Overseas Chinese Scholars, Ministry of
Education, and the Natural Science Foundation of Shandong Province
of P. R. China.
\appendix
\section{Receiver response speed and nonstationary probability density model}

In Eq.~(\ref{eq:five}), the Fokker-Planck operator
$L_{FP}=\frac{\partial}{\partial
x}V'(x)+\frac{D}{\tau_a}\frac{\partial^2}{\partial x^2}$ is not a
Hermitian operator \cite{Risken}. We rescale the variables as
\begin{eqnarray}
\bar X_b=X_b/ \sqrt{D/\tau_a}, \; \bar A_i =A_i/ \sqrt{D/\tau_a},
\nonumber \\ \tau=t/ \tau_a, \; y=x/ \sqrt{D/\tau_a},
\label{eq:A1}
\end{eqnarray}
Eq.~(\ref{eq:five}) becomes
\begin{equation}
\frac{\partial\rho(y,\tau)}{\partial
\tau}=\big[\frac{\partial}{\partial
y}V'(y)+\frac{\partial^2}{\partial y^2}\big] \rho(y,\tau),
\label{eq:A2}
\end{equation}
where $V'(y)=-(y-y^3/\bar X_b^2+\bar A_i)$. The steady-state
solution of Eq.~(\ref{eq:A2}) is given by
\begin{equation}
\rho(y)=\lim_{\tau\rightarrow\infty}\rho(y,\tau)=C\exp[-V(y)],\label{eq:A3}
\end{equation}
where $C$ is the normalization constant. A separation ansatz for
$\rho(y,\tau)$ \cite{Risken},
\begin{equation}
\rho(y,\tau)=u(y)\exp[-V(y)/2]\exp(-\lambda \tau), \label{eq:A4}
\end{equation}
leads to
\begin{equation}
L u=-\lambda u, \label{eq:A5}
\end{equation}
with a Hermitian operator $L=\frac{\partial^2}{\partial
y^2}-[\frac{1}{4}V'^2(y)-\frac{1}{2}V''(y)]$. The functions $u(y)$
are eigenfunctions of the operator $L$ with the eigenvalues
$\lambda$. Multiplying both sides of Eq.~(\ref{eq:A5}) by $u(y)$
and integrating it, yields
\begin{equation}
\lambda=\frac{\int_{-\infty}^{+\infty}\{u'^2(y)+u^2(y)[\frac{1}{4}V'^2(y)-\frac{1}{2}V''(y)]\}dy}
{\int_{-\infty}^{+\infty}u^2(y)dy}, \label{eq:A6}
\end{equation}
where eigenfunctions $u(y)$ satisfy the boundary conditions of
$u(y)$ and $u'(y)$ tending to zero as $y\rightarrow \pm\infty$.
The eigenvalue problem of Eq.~(\ref{eq:A5}) is then equivalent to
the variational problem consisting in finding the extremal values
of the right side of Eq.~(\ref{eq:A6}) \cite{Risken,Xu}. The
minimum of this expression is then the lowest eigenvalue
$\lambda_0=0$, corresponding to the steady-state solution of
Eq.~(\ref{eq:A3}) \cite{Risken}. We adopt here eigenfunctions
$u(y)=p(y)\exp[-V(y)/2]$ and $p(y)\neq 0$, Eq.~(\ref{eq:A6})
becomes
\begin{eqnarray}
\lambda=\{\int_{-\infty}^{+\infty}\{p'^2(y)+\frac{1}{2}p^2(y)V'^2(y)-\nonumber \\
\frac{1}{2}[V'(y)p^2(y)]'\}\exp[-V(y)]dy\}\nonumber \\
/\{\int_{-\infty}^{+\infty}p^2(y)\exp[-V(y)]dy\}. \label{eq:A7}
\end{eqnarray}
Since
\begin{eqnarray*}
\int_{-\infty}^{+\infty}[V'(y)p^2(y)]'\exp[-V(y)]dy\\=
V'(y)p^2(y)\exp[-V(y)]|_{-\infty}^{+\infty}\\
+\int_{-\infty}^{+\infty}p^2(y)V'^2(y)\exp[-V(y)]dy\\=\int_{-\infty}^{+\infty}p^2(y)V'^2(y)\exp[-V(y)]dy,
\end{eqnarray*}
Eq.~(\ref{eq:A7}) can be rewritten as
\begin{equation}
\lambda=\frac{\int_{-\infty}^{+\infty}p'^2(y)\exp[-V(y)]dy}
{\int_{-\infty}^{+\infty}p^2(y) \exp[-V(y)]dy}. \label{eq:A8}
\end{equation}
Assume $p(y)=d_0+d_1 y+\ldots +d_n y^n$ and the order $n$ is an
integer, we obtain
\begin{equation}
([K]-\lambda [M])\{d\}=0, \label{eq:A9}
\end{equation}
with eigenvectors $\{d^i\}=[d_0^i, d_1^i, \ldots d_n^i]$
corresponding to eigenvalues $\{\lambda\}=[\lambda_0,
\lambda_1,\ldots,\lambda_n]$ for $i=0,1,\ldots n$. The integer $n$
is not increased in the iterative process until the preceding
values of $\lambda_i$ approximate the next ones within the
tolerance error. The elements of matrices $[M]$ and $[K]$ are
\begin{eqnarray*}
m_{ij}=\int_{-\infty}^{+\infty}y^{i+j}\exp[-V(y)]dy>0,\\
k_{ij}=\int_{-\infty}^{+\infty} ij y^{i+j-2}\exp[-V(y)]dy\geq0,
\end{eqnarray*}
where $i,$ $j=0,1,\ldots n$. The matrix $[M]$ is positive definite
and the matrix $[K]$ is semi-positive definite. The minimal
eigenvalue $\lambda_0$ is zero. The minimal positive eigenvalue
$\lambda_1$ describes the main speed of the system tending to the
steady state solution of Eq.~(\ref{eq:A3}), what we call the
receiver response speed.

From Eq.~(\ref{eq:A9}), we can obtain the eigenfunctions
$u_i(y)=p_i(y)\exp[-V(y)/2]$ corresponding to the eigenvalue
$\lambda_i$ for $i=0,1,\ldots n$, where $p_i(y)=d_0^i+d_1^i y+
\dots +d_n^i y^n$. The eigenvectors $\{d^i\}=[d_0^i, d_1^i, \dots
d_n^i]$ are normalized. Because $L$ is a Hermitian operator,
eigenfunctions $u_i(y)$ and $u_j(y)$ are orthogonal
\begin{equation}
\int_{-\infty}^{+\infty}u_i(y)u_j(y)dy=\delta_{ij}, \label{eq:A10}
\end{equation}
where $i,j=0,1,\ldots,n$. Hence, $\rho(y,\tau)$ can be expanded,
according to eigenfunctions $u_i(y)$ and eigenvalues $\lambda_i$,
as
\begin{eqnarray*}
\rho(y,\tau)=\sum_{i=0}^n C_i u_i(y)\exp[-V(y)/2]
\exp[-\lambda_i\tau], \end{eqnarray*} where $C_{i}$ are
normalization constants deduced from the orthogonal condition of
eigenfunctions \cite{Risken}. Note the scale transformation in
Eq.~(\ref{eq:A1}), $\rho(x,t)$ can be represented as
\begin{equation}
\rho(x,t)=\sum_{i=0}^n C_i u_i(x)\exp[-\tau_a V(x)/(2D)]
\exp[-\lambda_i t], \label{eq:A11}
\end{equation}
with the real eigenvalues $\lambda_i=\lambda_i'/\tau_a$ in the
timescale of $t$ and $\lambda_i'$ derived from Eq.~(\ref{eq:A9}).

\end{document}